# Interfacial Spin-to-Charge Conversion in Sputtered $MoTe_2$ Heterostructures Probed by Spin Pumping and Spin-Torque Ferromagnetic Resonance

J. L. Costa[1], E. Santos[1], E. L. T. França[1], J. B. S. Mendes[2*], and A. Azevedo[1*]

[1]*Departamento de Física, Universidade Federal de Pernambuco, 50670-901, Recife, PE, Brazil.*

[2]*Departamento de Física, Universidade Federal de Viçosa, 36570-900, Viçosa, MG, Brazil.*

## ABSTRACT

Transition metal dichalcogenides (TMDs) and their Weyl semimetal phases, such as $MoTe_2$, have attracted significant attention for spin-orbit torque applications due to their efficient charge-to-spin conversion. However, whether this conversion originates predominantly from the bulk or the interface remains unclear. Here, we investigate spin-charge interconversion in $MoTe_2$ using spin-pumping and spin-torque ferromagnetic resonance (SP-FMR and ST-FMR). Thickness-dependent measurements reveal large spin-to-charge conversion and spin-torque efficiencies that are essentially independent of $MoTe_2$ thickness, indicating that the conversion is predominantly governed by the Rashba-Edelstein effect at the Py/$MoTe_2$ interface rather than by bulk spin transport. This behavior contrasts with the characteristic thickness dependence observed in Pt heterostructures and is further supported by bidirectional SP-FMR and interface-separation measurements. Our results highlight the dominant role of the Py/$MoTe_2$ interface in enabling efficient spin-charge conversion and spin-orbit torques in TMD-based spintronic devices. These findings highlight the potential of sputtered $MoTe_2$/Py heterostructures for low-power spintronic applications, including magnetic memory and logic devices.

[*]**Corresponding authors:** J. B. S. Mendes and A. Azevedo.

**E-mail:** joaquim.mendes@ufv.br, antonio.azevedo@ufpe.br

## 1. INTRODUCTION

Since the discovery of two-dimensional (2D) materials, their unique properties have motivated extensive research in both fundamental physics and device applications[1-5]. Among these, transition metal dichalcogenides (TMDs) have emerged as particularly promising platforms due to their strong spin-orbit coupling (SOC), tunable band structure, and reduced crystalline symmetry[6-9]. These properties have positioned TMDs as leading candidates for spintronic devices requiring lower power consumption and efficient charge-spin interconversion[10-14].

Recent research has focused on low-symmetry TMDs such as $MoTe_2$ and $WTe_2$, where broken inversion symmetry combined with strong SOC gives rise to interfacial Rashba interactions and spin momentum-locking[15-23]. Interfaces with ferromagnetic materials can further enhance these effects through orbital hybridization and increased structural asymmetry. In this framework, the perpendicular electric field at the interface, together with the strong SOC, produces an effective magnetic field acting on conduction electrons. As a result, the spin orientation becomes locked to the momentum. This interaction can be interpreted as an effective Zeeman-like term that lifts the spin degeneracy of the electronic bands and, under an applied current, generates a non-equilibrium spin accumulation at the interface[24]. This current-induced spin accumulation is known as the spin Rashba-Edelstein effect (SREE), which converts charge currents into spin accumulation, while its reciprocal process, the inverse SREE (ISREE), converts spin current into charge current[24,25]. These interface-driven mechanisms are fundamentally different from conventional spin Hall effects, in which the charge-to-spin interconversion occurs within the bulk and without spin-momentum locking.

Among TMDs, the identification of materials that combine strong SOC effects with compatibility with scalable fabrication techniques is of particular interest for spintronics applications. $MoTe_2$ stands out among TMDs due to its polymorphic phases (2H, 1T', Td), temperature- and strain-dependent topological transitions, and large reported spin-torque efficiencies, even for sputtered samples[15-21, 26-28]. For example, $MoTe_2$-based heterostructures have demonstrated larger spin-torques efficiencies than those reported for heavy metals with strong SOC, such as Pt, Ta and W[29-31]. Additionally, $MoTe_2$ has been reported to exhibit unconventional spin-orbit torques in some heterostructures, a phenomenon associated with broken mirror symmetry and therefore not generally expected in polycrystalline samples[32-34]. The observation of such effects in sputtered $MoTe_2$-based systems is particularly relevant because it suggests that high spin-orbit torque efficiencies can be achieved without relying exclusively on growth techniques that produce high-quality single-crystalline TMDs, such as CVD and MBE[35-37]. From both scientific and technological perspective, these results highlight sputter deposition as a promising route toward the scalable fabrication of spintronics devices.

Despite the high efficiencies reported for $MoTe_2$, the relative contributions of bulk versus interfacial mechanisms remain a subject of intense debate. This is particularly true for polycrystalline sputtered films, where the conversion efficiency is highly sensitive to growth parameters, substrate choice, and phase stoichiometry[33,34,38]. In this work, we address this issue through a comparative study of sputtered $MoTe_2$ and Pt heterostructures using spin pumping driven by FMR SP-FMR) and spin-torque ferromagnetic resonance (ST-FMR). We first investigate spin-to-charge interconversion by SP-FMR and find that the conversion efficiency in $MoTe_2$ remains nearly independent of thickness, in strong contrast to the bulk-dominated behavior observed in Pt reference samples. This interpretation is further supported by bidirectional spin pumping measurements, which provide a direct means to distinguish between bulk and interface conversion mechanisms. We then investigate the reciprocal charge-to-spin conversion using ST-FMR. Angular-dependent measurements reveal signatures of unconventional out-of-plane torques associated with symmetry-breaking interfacial effects. By combining SP-FMR, ST-FMR, and interface-engineering experiments, our results demonstrate that the dominant conversion mechanism in these heterostructures is governed primarily by Rashba-type interfacial effects rather than bulk spin Hall transport.

## 2. EXPERIMENTAL RESULTS AND THEORETICAL DISCUSSION

### 2.1 SP-FMR in $SiO_2$/Py/$MoTe_2$ heterostructures

To investigate the spin-to-charge conversion in $MoTe_2$, we initially employed the SP-FMR technique, which allows us to detect the inverse spin Hall effect (ISHE) or ISREE[24,25,39-45]. The detailed experimental setup and parameters are explained in appendix B. Appendix A shows the characterization of our $MoTe_2$ films. In the SP-FMR, the magnetization precession of a ferromagnetic (FM) layer transports spin angular momentum to the adjacent layer (generally non-magnetic, NM), creating a non-equilibrium spin accumulation that undergoes diffusion across the bulk of the adjacent material, thereby generating a spin current[34-38]. The spin current can be converted into a measurable charge current in the transversal direction, either through the bulk of the NM layer or through the interface NM/FM between the two layers, as described by the equations

$$\vec{J}_C^{ISHE} = \left(\frac{2e}{\hbar}\right)\theta_{SH}\left(\hat{\sigma}_S \times \vec{J}_S\right), \quad \text{bulk interconversion} \tag{1.1}$$

$$\vec{J}_C^{IREE} = \left(\frac{e\alpha_R}{\hbar}\right)\left(\hat{z} \times \vec{S}_{neq}\right), \quad \text{interface interconversion} \tag{1.2}$$

where $\hat{\sigma}_S$ is the spin polarization of the spin current injected by the FM layer, $\theta_{SH}$ is the spin Hall angle, $e$ is the elementary charge and $\vec{S}_{neq}$ is the non-equilibrium spin density, which is polarized perpendicular to the charge current direction. In our experiments, the bulk interconversion corresponds to the ISHE, while the interface interconversion corresponds to the ISREE.

The investigation of the spin-to-charge conversion in $MoTe_2$ by the spin-pumping technique was initially done using two types of magnetic materials, YIG ($Y_3Fe_5O_{12}$) and Py ($Ni_{81}Fe_{19}$). Despite the ferrimagnet YIG ($Y_3Fe_5O_{12}$) being the standard material for spin pumping experiments in ferromagnet/non-magnet (FM/NM) heterostructures, we were unable to obtain consistent SP-FMR signals in YIG/$MoTe_2$ bilayers, in agreement with previous reports on this system[46]. This behavior is likely related to the inefficient spin injection from YIG into two-dimensional van der Waals materials, which has also been observed in other YIG/2D heterostructures. Therefore we employed Py as the spin injector. Although Py-based spin pumping experiments are accompanied by additional contributions, such as galvanomagnetic effects (anomalous Hall effect, anisotropic magnetoresistance, etc.) and spin-to-charge self-conversion within the ferromagnetic layer[43,47-50], Py provides a robust and reliable source of spin current. Accordingly, spin pumping measurements were performed on a series of $SiO_2$/$MoTe_2$(t)/Py(7) samples, where the 7-nm-thick Py layer serves both as the spin injector and as a protective capping layer that prevents the oxidation of $MoTe_2$. Fig. 1 (a) schematically illustrates the SP-FMR geometry employed in this work. Under ferromagnetic resonance, a spin current $\vec{J}_S$ is pumped from the Py layer into the underlying $MoTe_2$ layer along z-direction. The spin polarization $\hat{\sigma}_S$, which follows the magnetization direction, lies in the film plane, so that the spin-to-charge conversion generates a transverse charge current, $\vec{J}_C \propto \hat{\sigma}_S \times \vec{J}_S$, flowing along y direction. Figs. 1(b)-(d) show the SP-FMR signals measured for the $SiO_2$/$MoTe_2$(3)/Py(7) sample with the magnetic field applied at $\phi = 0°, 180°,$ and 90°, respectively. To directly quantify the spin-to-charge conversion, the electrical signal is expressed as the charge current $I_C = V/R$, where $V$ is the voltage measured by a nanovoltmeter and $R$ is the electrical resistance between the voltage probes. The solid black curves are Lorentzian fits to the experimental data. Reversing the magnetic field from $\phi = 0°$ [Fig.1(b)] to $\phi = 180°$ [Fig.1(c)] reverses the spin polarization, and consequently the sign of the converted charge current, as expected from $\vec{J}_C \propto \hat{\sigma}_S \times \vec{J}_S$. In contrast, for $\phi = 90°$ [Fig.1(d)], the charge current along the y direction is expected to vanish because $\hat{\sigma}_S \times \vec{J}_S$ is perpendicular to the voltage probes, in agreement with the negligible measured signal. Since ferromagnetic resonance in Py also produces intrinsic electrical signals, including galvanomagnetic effects and spin-to-charge self-conversion, the measured current contains both the contribution associated with $MoTe_2$ and the intrinsic response of the Py layer. Therefore, the measured current can be modeled as the sum of symmetric and antisymmetric Lorentzian functions,

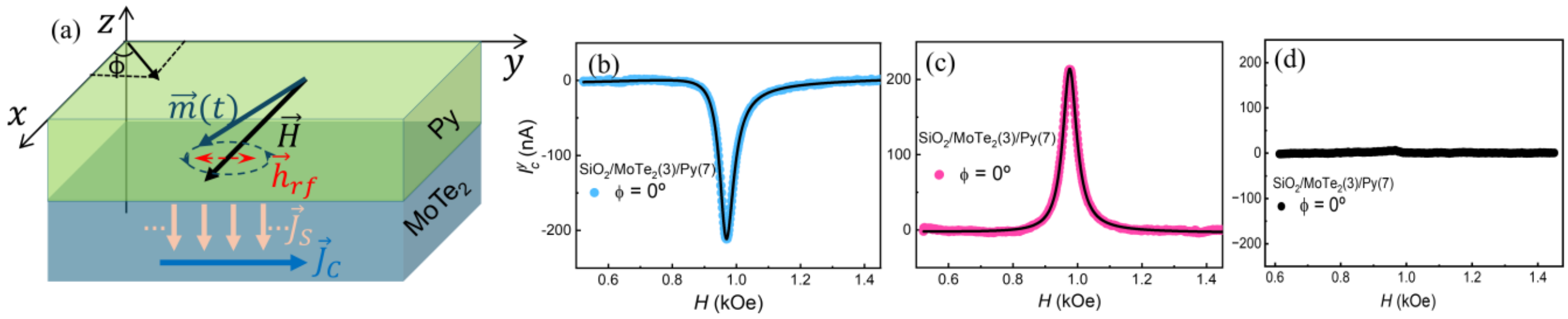

**Figure 1.** (a) Schematic illustration of the SP-FMR experimental geometry, showing the heterostructure stack and the detected signal. (b - d) Measured SP-FMR signal as a function of the in-plane applied magnetic field for the azimuthal angles $\phi = 0°$, $\phi = 180°$ and $\phi = 90°$, defined as the angle between the magnetic field and the current direction. The solid lines correspond to the best fits to the data using Eq. (2). The fits yield $I_S$ = -216.8 nA ($\phi = 0°$), $I_S$ = 213.9 nA ($\phi = 180°$), confirming the spin-pumping origin, while $I_A$ =-46.1 nA and $I_A$ =10.7 nA, respectively. At $\phi = 90°$ , the symmetric signal vanishes as expected. All measurements were performed at an applied RF power of 110 mW.

$$I = I_S \frac{\Delta H^2}{[(H - H_r)^2 + \Delta H^2]} + I_A \frac{\Delta H(H - H_r)}{[(H - H_r)^2 + \Delta H^2]}, \quad (2)$$

where $H_r$ is the resonance field, $\Delta H$ is the half-width at half-maximum, and $I_S$ and $I_A$ are the amplitudes of the symmetric and antisymmetric Lorentzian components, respectively. The sign of the fitted signal follows the direction of the spin-to-charge conversion predicted by Eq. 1.1 or 1.2, providing a direct consistency check with the expected polarity of the bulk ISHE and the interfacial ISREE.

To investigate the spin-to-charge conversion mechanisms in Py/$MoTe_2$ bilayers, we systematically studied the dependence of the SP-FMR signals on the $MoTe_2$ thickness by fabricating $SiO_2$/$MoTe_2$($t_{MoTe_2}$)/Py(7) heterostructures with $t_{MoTe_2} = 3$, 4, 8, 10, 15, and 20 nm. To account for the self-conversion signal of the Py layer, a 7 nm Py reference sample was also measured under identical conditions. The corresponding SP-FMR spectra are presented in Figs. 2(a)-(h), together with the fitting curves obtained using Eq. (2). Measurements were performed for two opposite magnetic-field orientations ($\phi = 0°$, and $\phi = 180°$), allowing the separation of the spin-pumping contribution from spurious voltage signals. As expected for a genuine spin-pumping induced charge current, reversing the magnetic field orientation results in a complete inversion of the signal polarity while preserving both the resonance field and the spectral lineshape. The excellent agreement between the experimental data and the fitting curves confirms that the spectra are well described by the adopted fitting model.

A comparison between the reference Py film [Fig. 2(a)] and the Py/$MoTe_2$ bilayers [Fig. 2(b)-(h)] reveals a pronounced enhancement of the detected current upon the introduction of the $MoTe_2$ layer. While the self-conversion current of Py reference is approximately $I_S$=110 nA, the heterostructures exhibit currents ranging from about 190 to 230 nA, nearly doubling the signal intensity. This enhancement clearly indicates an additional spin-to-charge conversion contribution associated with the $MoTe_2$ layer. Furthermore, despite the $MoTe_2$ thickness varying from 3 to 20 nm, the signal amplitude remains nearly constant, exhibiting only moderate sample-to-sample variations without any systematic thickness dependence. This behavior suggests that the dominant conversion mechanism is largely thickness independent over the investigated range, consistent with a process governed primarily by the interface or by a characteristic length shorter than the investigated thickness range. Additionally, both the resonance field and linewidth remain essentially unchanged for all samples, indicating that the magnetic properties of the 7 nm Py layer are only weakly affected by the presence and thickness of the $MoTe_2$ layer.

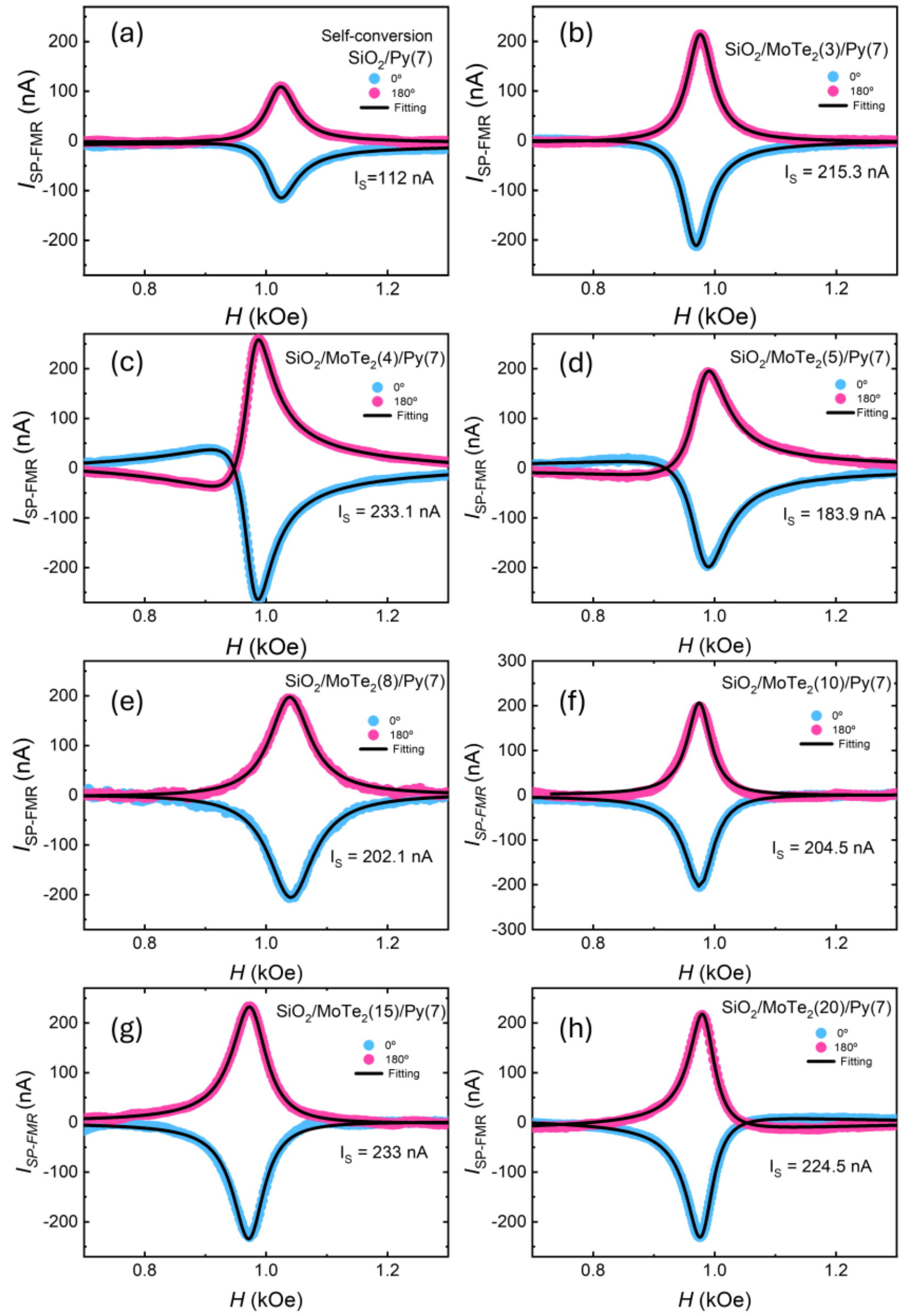

**Figure 2**. SP-FMR spectra measured for the reference SiO2/Py(7) film (a) and SiO2/MoTe2($t_{MoTe_2}$)/Py(7) bilayers with $t_{MoTe_2}$= 3 (b), 3 (c), 5 (d), 8 (e), 10 (f), 15 (g) and 20 nm (h). The blue and pink symbols correspond to measurements with $\phi = 0°$ and $\phi = 180°$, respectively. The solid black curves are fits to Eq. 2. The measurements were carried out at an rf frequency of 9.4 GHz and an rf power of 110 mW.

To better visualize these results, the magnitude of the symmetric SP-FMR components ($I_S$) is plotted as a function of MoTe2 thickness in Fig. 3(a). The values of $I_S$ were obtained by averaging the absolute values extracted from the measurements performed with $\phi = 0°$ and $\phi = 180°$. The error bars include both the fitting uncertainty and variance between the two field orientations. For comparison, analogous measurements were carried out on SiO2/Pt($t_{Pt}$)/Py(7) bilayers, which exhibits the thickness dependence expected for a bulk spin-to-charge conversion process [Fig. 3(b)]. The complete SP-FMR data for the Pt/Py samples are presented in Fig. A.3 of Appendix C. Fitting the data in Fig. 3(b) with $I_{Peak} \propto \tanh(t_{Pt}/2\lambda_S)$ yields a spin diffusion length of $\lambda_S = (1.1 \pm 0.1)$ nm. The data point at $t_{Pt} = 0$ corresponds to the self-conversion signal of the Py(7) reference layer. In contrast, the SP-FMR signal measured in the MoTe2/Py bilayer remains nearly constant over the entire investigated thickness range, showing only small sample-to-sample variations around an average value of approximately 210 nA, well above the Py self-conversion signal. The absence of a systematic thickness dependence is consistent with a spin-to-charge conversion process dominated by the $MoTe_2$/Py interface, as expected for the ISREE.

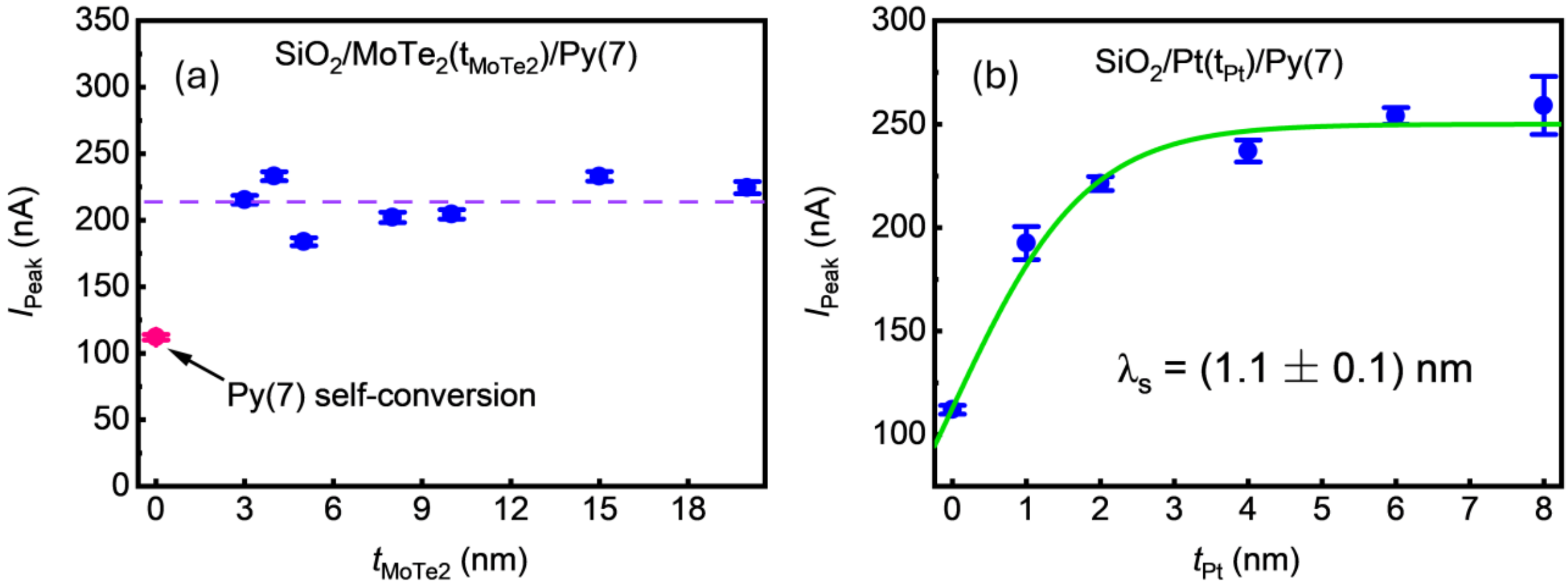

**Figure 3.** Dependence of the SP-FMR peak current ($I_{Peak}$) on the thickness of the spin-to-charge conversion layer. (a) $SiO_2/MoTe_2(t_{MoTe2})/Py(7)$ heterostructures exhibit an approximately thickness-independent response, consistent with an interface-dominated spin-to-charge conversion process. The dashed horizontal line indicates the average $I_{Peak}$, and the red symbol at $t_{MoTe2}$=0 represents the self-conversion signal Py(7) reference film. (b) $SiO_2$/Pt ($t_{Pt}$)/Py(7) heterostructures display the characteristic thickness dependence expected for a bulk spin-diffusion process. The solid line is a fit to $I_{Peak} \propto tanh(t_{Pt}/2\lambda_S)$, yielding a spin diffusion length of $\lambda_S = (1.1 \pm 0.1)$ nm.

To further verify the interfacial origin of the spin-to-charge conversion inferred from the thickness-independent SP-FMR amplitudes, we performed SP-FMR measurements in a trilayer geometry that enables spin-current injection into the nonmagnetic layer from opposite directions. This configuration is realized in the Co/$MoTe_2$/Py heterostructure, where spin pumping from either the Py or Co layer injects spin current into the $MoTe_2$ layer. For Rashba-type interfacial conversion, described by Eq. (1.2), the generated charge current is independent of the spin-current injection direction[51-55]. Consequently, spin currents injected from opposite sides produce electrical signals with the same polarity. As a control experiment, we also investigated Co/Pt/Py, where the dominant bulk spin-to-charge conversion in Pt is expected to follow Eq. (1.1), leading to opposite signal polarities for opposite spin-current injection directions. Fig. 4 summarizes the results obtained for Co(7)/Pt(10)/Py(7) [Figs. 4(a-c)] and Co(7)/$MoTe_2$(3)/Py(7) [(Figs. 4(d-f)). Since the measured SP-FMR signal contains contributions from both the spin-to-charge conversion in the nonmagnetic layer and the self-conversion in the ferromagnetic layers, the latter was subtracted using independent measurements on single Co and Py reference samples, as shown in the insets of Fig. 4(b). Two main conclusions emerge from these measurements. First, the spin-to-charge conversion at the Co/Pt and Co/$MoTe_2$ interfaces exhibit a relatively weaker spin-to-charge conversion than the Co/Py and $MoTe_2$/Py interfaces is weaker than at the Pt/Py and $MoTe_2$/Py interfaces, as evidenced by the smaller $I_{Peak}$ amplitudes. This behavior is likely associated with a lower spin-mixing conductance at the Co interface, possibly resulting from the larger roughness of the sputtered Co films. Second, in the Co/Pt/Py trilayer the two resonances exhibit opposite polarities, as expected for bulk spin-to-charge governed by Eq. 1.1. In contrast, in the Co/$MoTe_2$/Py trilayer both resonances have the same polarity, despite the opposite spin-current injection directions. This behavior is precisely that predicted by Eq. (1.2) for the ISREE, providing direct evidence that spin-to-charge conversion in $MoTe_2$ is dominated by interfacial Rashba states rather than by bulk states.

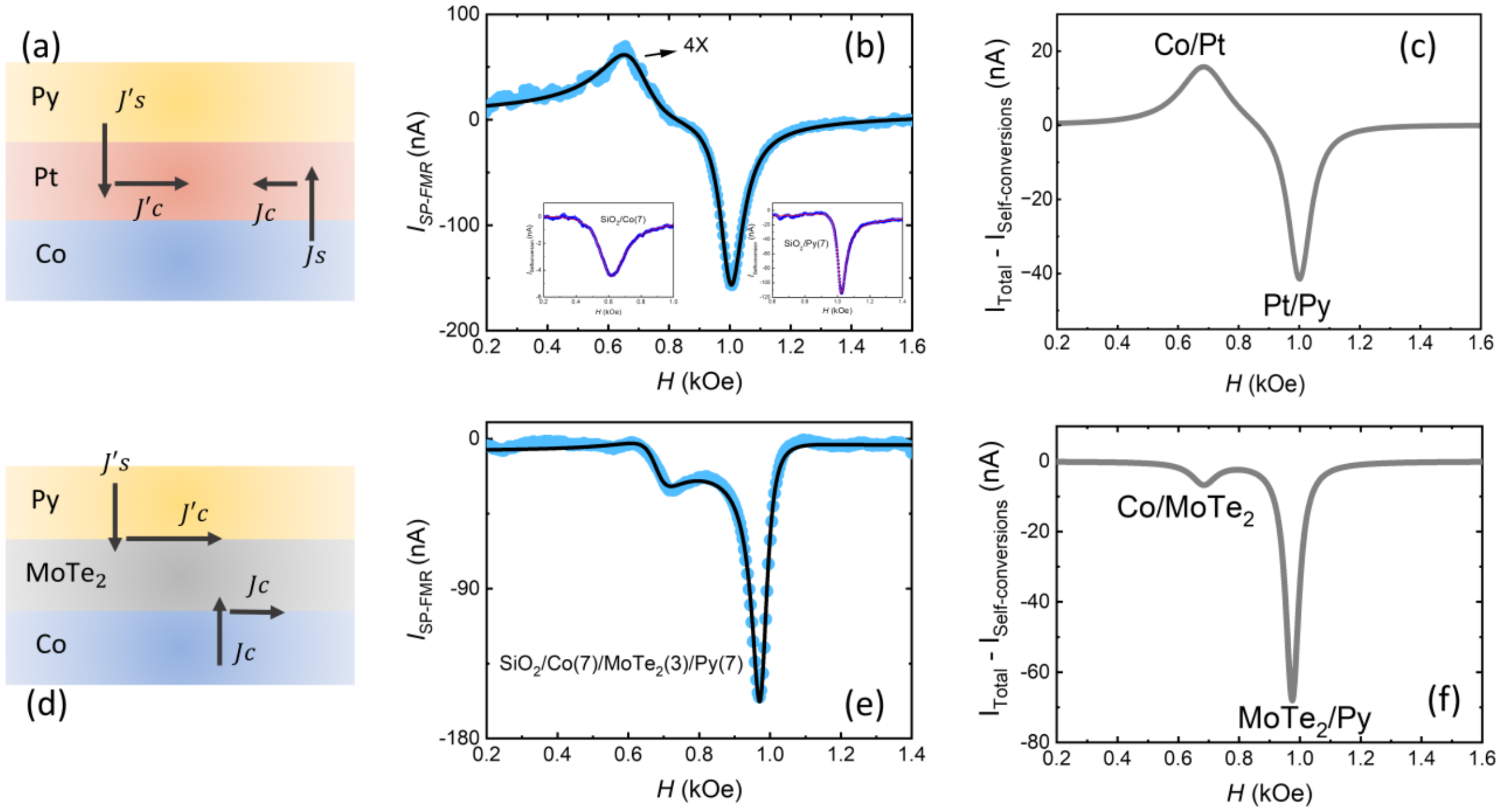

**Figure 4.** (a, d) Schematic illustration of the $SiO_2$/Co(7)/$MoTe_2$(3)/Py(7) and $SiO_2$/Co(7)/Pt(10)/Py(7) heterostructures, respectively, indicating the directions of charge current generated by the (a) ISHE and (d) ISREE. (b, e) SP-FMR spectra of the corresponding trilayers. The insets show the self-conversion signals measured independently in the Co(7) and Py(7) reference samples. (c, f) Spin-to-charge conversion signals obtained after subtracting the Co and Py self-conversion contributions from the total SP-FMR spectra in (b) and (e), respectively, thereby isolating the contributions from Pt and $MoTe_2$. All measurements were performed with an rf power of 110 mW.

### 2.2 ST-FMR in $SiO_2$/Py/$MoTe_2$ heterostructures

Fig. 5 shows a schematic representation of the ST-FMR experimental setup and the measured signals. As illustrated in Fig. 5(a), an rf current, $I_{rf}$, is applied to a FM/NM bilayer, where the NM in the present work is $MoTe_2$. The current flowing through the NM layer generates an rf Oersted field ($\vec{H}_{oe}$) that drives the magnetization dynamics of the adjacent FM layer. When the excitation frequency matches the natural precession frequency of the FM magnetization, the system reaches FMR condition. Simultaneously, $I_{rf}$ can be converted into a spin current ($\vec{J}_s$) either within the NM layer via the SHE or at the FM/NM interface through the spin Rashba effect (SREE), generating additional spin-orbit torques that modify the magnetization dynamics. These torques can be incorporated into the Landau-Lifshitz-Gilbert equation[56-61]

$$\frac{d\hat{m}}{dt} = -\gamma \hat{m} \times \vec{H}_{eff} + \alpha \left(\hat{m} \times \frac{d\hat{m}}{dt}\right) + \frac{\gamma}{M_s}(\vec{\tau}_{FL} + \vec{\tau}_{DL}), \quad (3)$$

where $\gamma$ is the gyromagnetic ratio, $\alpha$ is the gilbert damping parameter, $\vec{H}_{eff} = -\vec{\nabla}_M E$ is the effective magnetic field, and $\vec{\tau}_{FL} = \tau_{FL}(\hat{m} \times \hat{\sigma})$ and $\vec{\tau}_{DL} = \tau_{DL}[\hat{m} \times (\hat{m} \times \hat{\sigma})]$ are the field-like (FL) and damping-like (DL) spin-orbit torques, respectively, with $\hat{\sigma}$ denoting the spin current polarization direction. As illustrated in Fig. 5(a), $\vec{\tau}_{FL}$ is normal to the plane defined by $\hat{m}$ and $\hat{\sigma}$. It has the same symmetry as an effective magnetic field acting on the magnetization and therefore contributes to its precessional motion. In contrast, the damping-like torque lies within the plane defined by $\hat{m}$ and $\hat{\sigma}$, and has the same symmetry as the Gilbert damping torque. Consequently, it can either increase or decrease the effective magnetic damping, depending on its sign.

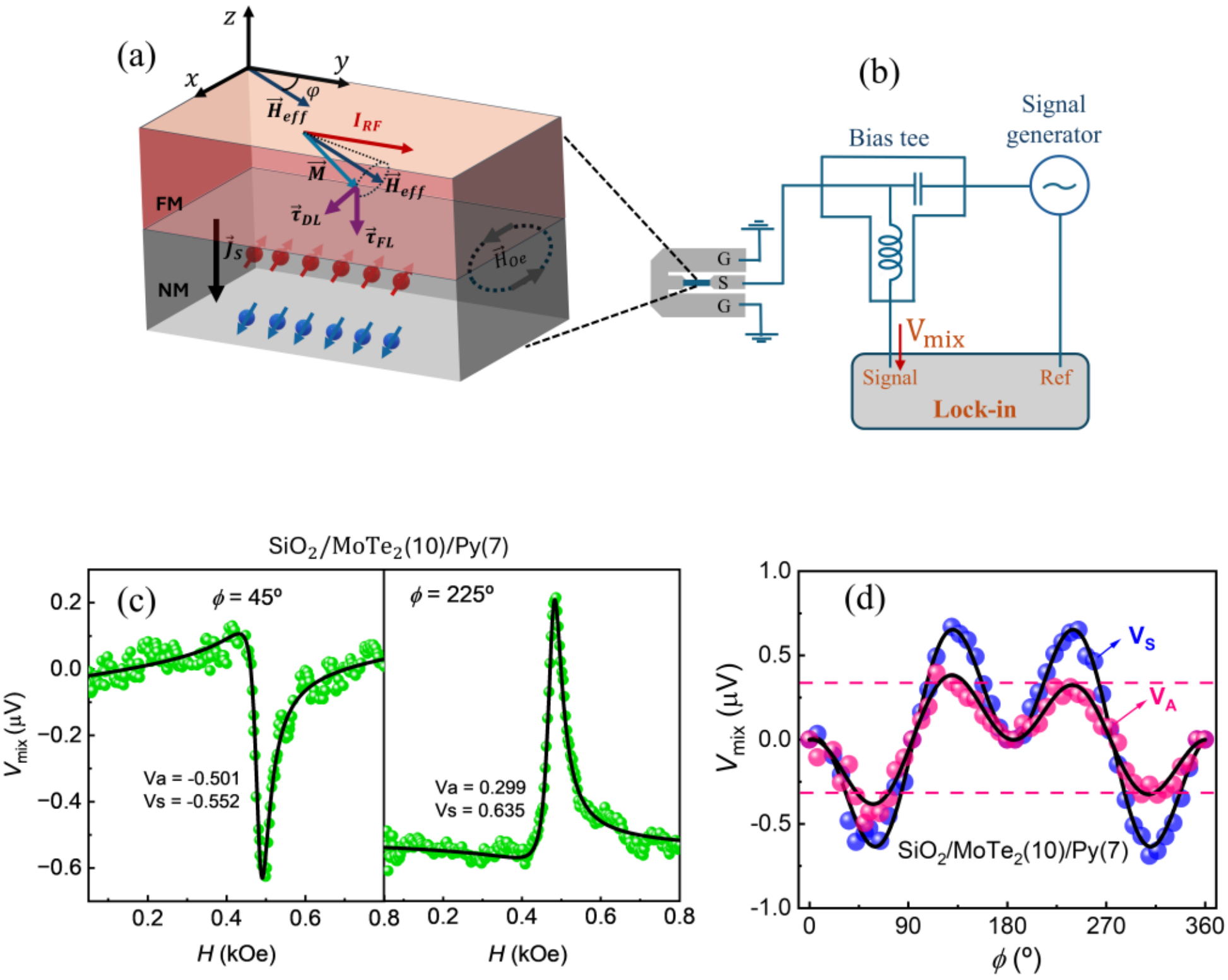

**Figure 5**. (a) Schematic illustration of the ST-FMR geometry for the $SiO_2$/NM/Py bilayer. An rf current $I_{rf}$ flowing through to the NM layer generates an Oersted field ($\vec{H}_{oe}$) and a spin current $\vec{J}_S$ via the bulk spin Hall effect or the interfacial Rashba-Edelstein effect (REE). The spin current exerts damping-like ($\vec{\tau}_{DL}$) and field-like ($\vec{\tau}_{FL}$) torques on the magnetization $\hat{m}$. (b) Schematic of the ST-FMR setup with lock-in detection used to measure the mixing voltage ($V_{mix}$) attributed to the combined action of $\vec{\tau}_{DL}$ and $\vec{\tau}_{FL}$. (c) Representative ST-FMR signal measured in $SiO_2$/$MoTe_2$(10)/Py(7) measured for opposite magnetic field directions. (d) Angular dependence of the symmetric ($V_S$) and antisymmetric ($V_A$) components of $V_{mix}$. Measurements were performed at a microwave frequency of 6.0 GHz and an rf power of 8.0 mW.

Since FM materials exhibit anisotropic magnetoresistance (AMR), magnetization precession under FMR produces an oscillating electrical resistance. The mixing of this time-dependent resistance with $I_{rf}$ generates a rectified voltage, detected as the ST-FMR signal using the lock-in technique illustrated in Fig. 5(b). The damping-like spin-orbit torque typically gives rise to a symmetric Lorentzian contribution, whereas the Oersted field and field-like torque mainly contribute to an antisymmetric Lorentzian component. Consequently, the measured ST-FMR spectrum consists of a superposition of symmetric and antisymmetric Lorentzian line shapes and can be described by[56-61]

$$V_{mix} = \; V_S \frac{\Delta H^2}{[(H - H_r)^2 + \Delta H^2]} + V_A \frac{\Delta H(H - H_r)}{[(H - H_r)^2 + \Delta H^2]}, \tag{4}$$

where $H_r$ is the resonance field, $\Delta H$ is the half-width at half-maximum, and $V_S$ and $V_A$ are the amplitudes of the symmetric and antisymmetric Lorentzian components, respectively. A representative ST-FMR signal for the $SiO_2$/$MoTe_2$/Py sample is shown in Fig. 5(c) (green symbols). The dark solid curves represent the addition of symmetric $V_S$ and antisymmetric $V_A$ components of the signal, obtained by fitting the experimental data using Eq. (4), with the corresponding fitted values indicated in Fig. 4c). Notably, $V_A$ exhibits an asymmetry under a 180∘ rotation of the magnetic field: its magnitude differs significantly between $\phi = 45°$ and 225°. This behavior is not accounted for by the conventional angular dependence shown by the solid curve in Fig. 5(d), indicating the presence of an additional contribution to the ST-FMR signal, as discussed below.

The extracted values of $V_S$ and $V_A$ are essential for determining the effective spin-torque efficiency, ($\xi_{FMR}$), which is calculated from [56-61]

$$\xi_{FMR} = \frac{V_S}{V_A}\frac{e}{\hbar}\mu_0 M_S t_{FM} t_{NM}\sqrt{1+\frac{4\pi M_{eff}}{H_r}}, \quad (4.1)$$

where $M_S$ is the saturation magnetization, $t_{NM}$ and $t_{FM}$ are the thicknesses of the NM and FM layers, respectively, and $4\pi M_{eff}$ is the effective magnetization determined from the Kittel equation $\omega = \gamma\sqrt{H(H+4\pi M_{eff})}$. Assuming that the damping-like and field-like torque efficiencies, $\xi_{DL}$ and $\xi_{FL}$, independent of the FM layer thickness, they can be extracted from the linear dependence of $1/\xi_{FMR}$ as a function of $1/t_{FM}$[56-61],

$$\frac{1}{\xi_{FMR}} = \frac{1}{\xi_{DL}}\left(1+\frac{\hbar}{e}\frac{\xi_{FL}}{\mu_0 M_s t_{NM} t_{FM}}\right). \quad (4.2)$$

The thickness dependence of $\xi_{FMR}$ can be used to qualitatively assess whether the generated spin torque originates predominantly from bulk or interfacial mechanisms. For torques generated in the NM bulk via SHE, $\xi_{FMR}$ is expected to increase with NM thickness and eventually saturate once the thickness exceeds the spin diffusion length. In contrast, for torques generated predominantly by interfacial mechanisms, such as the Rashba-Edelstein effect, $\xi_{FMR}$ is expected to remain nearly constant with increasing NM thickness. To investigate this behavior, we fabricated two sets of samples. In the first set, the FM layer thickness was varied while keeping the $MoTe_2$ thickness constant, enabling a detailed analysis of the DL and FL torque efficiencies. In the second set, the $MoTe_2$ thickness was varied while keeping the magnetic layer thickness fixed, allowing us to evaluate whether the spin-to-charge interconversion originates predominantly from the $MoTe_2$ bulk or from the FM/$MoTe_2$ interface.

To corroborate the thickness-dependent SP-FMR results, we performed ST-FMR measurements on $SiO_2$/$MoTe_2$($t_{\mathrm{MoTe_2}}$)/Py(7) heterostructures with $t_{\mathrm{MoTe_2}}$ = 2 - 20 nm. The Py layer was deposited on top of the $MoTe_2$ layer to protect it from oxidation. In addition to field-swept measurements, angle-dependent ST-FMR experiments were carried out by varying the in-plane angle $\phi$ from 0 to 360º, where $\phi$ is the angle between the external magnetic field $H$ and the rf current $I_{rf}$. Such measurements are essential for identifying unconventional torque components that cause angular asymmetries, as previously reported[32, 31, 61].

Fig. 6 summarizes the experimental measurements. The left column shows the ST-FMR spectra ($V_{\mathrm{mix}}$) for fixed angles of $\phi = 225°$ and $\phi = 45°$. The spectra exhibit pronounced symmetric Lorentzian contributions, particularly for thinner $MoTe_2$ layers. As the $MoTe_2$ thickness increases, the antisymmetric component becomes progressively larger, mainly because a larger fraction of the rf current flows through the $MoTe_2$ layer, thereby enhancing the Oersted field. In addition to these measurements, we performed in-plane angular measurements, as shown in. The right column of Fig. 6 shows the angular dependences of symmetric ($V_S$, blue symbols) and antisymmetric ($V_A$, pink symbols) components extracted from the ST-FMR spectra. Clear angular asymmetries are observed in both $V_S$ and $V_S$, which cannot be explained solely by the conventional torque components. For example, the absolute value of $V_S$ (and $V_A$) at $\phi = 45°$ differs from that at $\phi = 225°$, demonstrating that the response is not symmetric under a $180°$ rotation. This behavior is reproduced by the fitting curves and reveals the presence of an additional out-of-plane torque component.

Phenomenologically, in addition to the conventional $\cos\phi \sin 2\phi$ dependence of $V_S$ and $V_A$, the experimental data require an additional $\sin 2\phi$ term. This contribution naturally arises by writing the damping-like (DL) and field-like (FL) torques as $\tau_{DL}(\phi) = \tau_{DL}^{0} \cos\phi + \tau_{DL}^{\perp}$, and $\tau_{FL}(\phi) = \tau_{FL}^{0} \cos\phi + \tau_{FL}^{\perp}$, where $\tau_{DL}^{\perp}$ and $\tau_{FL}^{\perp}$ are angle-independent out-of-plane contributions. Since $V_S$ and $V_A$ are proportional to $\tau_{DL}(\phi)$ and $\tau_{FL}(\phi)$, respectively, and also $dR/d\phi$, with the AMR resistance following $R \propto \cos^2\phi$, the angular dependence can be written as[29,32,58,59,61,62]

$$V_S(\phi) \propto (\tau_{DL}^{0} \cos\phi + \tau_{DL}^{\perp}) \sin 2\phi, \tag{5.1}$$

$$V_A(\phi) \propto (\tau_{FL}^{0} \cos\phi + \tau_{FL}^{\perp}) \sin 2\phi. \tag{5.2}$$

The observation of out-of-plane torque components indicates an additional symmetry breaking beyond that expected for ideal isotropic metallic bilayers. Such asymmetries are typically attributed to mirror symmetry breaking, structural inversion asymmetry, or the generation of spin currents with out-of-plane spin polarization[29,32,58,59,61,62]. In the present $MoTe_2$ system, this symmetry reduction likely originates at the FM/$MoTe_2$ interface, where the structural inversion asymmetry, interfacial strain, or local disorder can break the spatial symmetry thereby leading to the generation of a spin current with out-of-plane polarization.

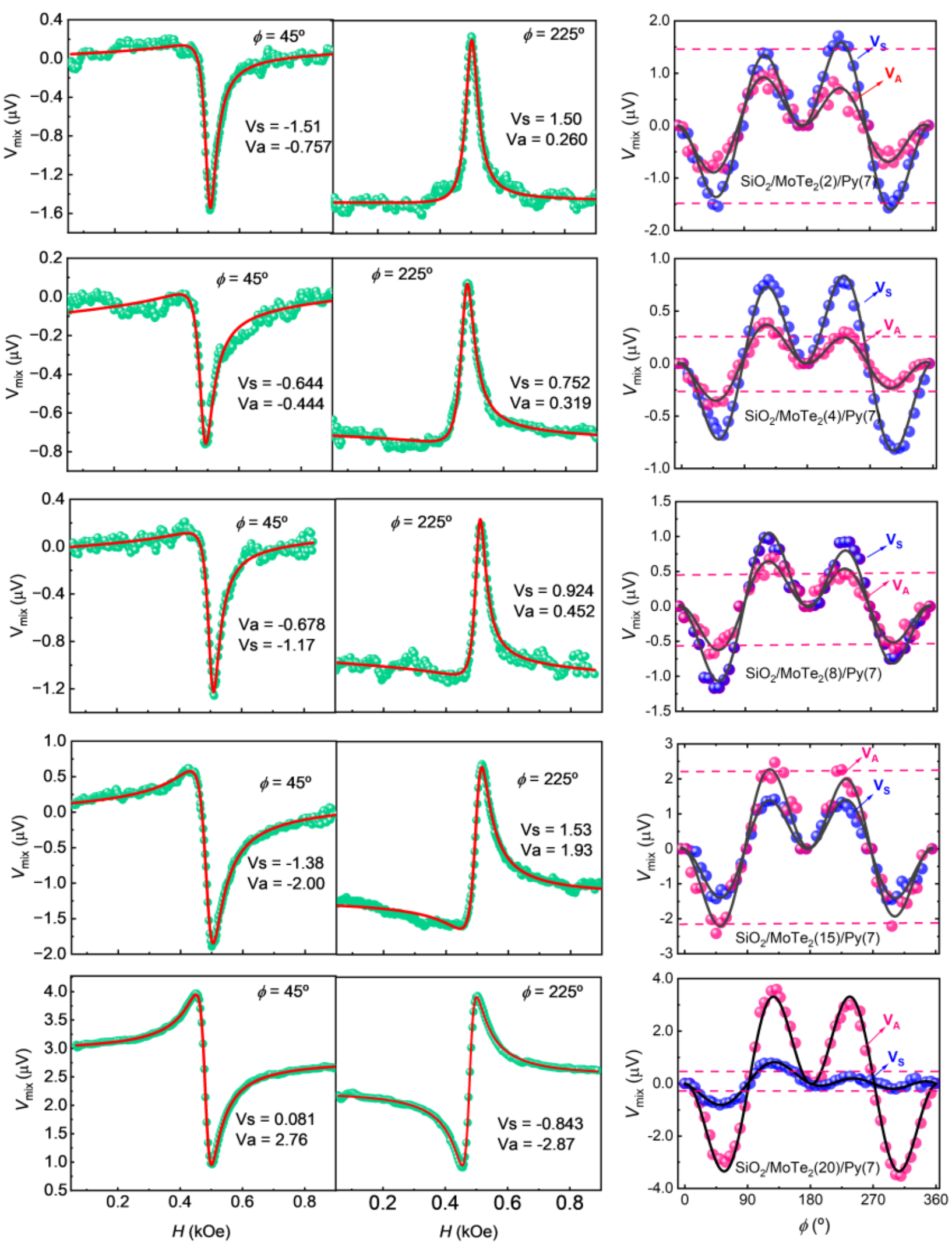

**Figure 6.** ST-FMR measurements for $SiO_2/MoTe_2(t_{MoTe2})/Py(7)$ heterostructures with $MoTe_2$ = 2, 4, 8, 15 e 20 nm. Left panels: $V_{mix}$ signals acquired at $\phi = 225^o$ and $\phi = 45^o$, fitted with symmetric ($V_S$) and antisymmetric ($V_A$) Lorentzian components. Right panels: angular dependence of $V_S$ and $V_A$ components, fitted using Eqs. (4.1) and (4.2). The dashed horizontal lines highlight the angular asymmetries. All measurements were performed at a microwave frequency of 6.0 GHz and an RF power of 8.0 mW.

To further distinguish between bulk and interface charge-spin conversion, we used Eq. 4.1 to calculate the parameter $\xi_{FMR}$, for the $SiO_2/MoTe_2(t_{MoTe_2})/Py(7)$ bilayers and compared the results with those obtained for the control $SiO_2/Pt(t_{Pt})/Py(7)$ heterostructure. For each sample $\xi_{FMR}$ was calculated using the fitted amplitudes at the angle corresponding to maximum ST-FMR signal ($\phi = 45$º). Fig. 7(a) summarizes the results. The black symbols correspond to the $MoTe_2$/Py bilayers, whereas the red symbols represent the Pt/Py samples (multiplied by a factor of two for clarity). In the $MoTe_2$/Py heterostructures, $\xi_{FMR}$ remains nearly independent of the $MoTe_2$ thickness, fluctuating around an average value. In contrast, the Pt/Py samples exhibit the expected asymptotic increase with Pt thickness, followed by saturation at the spin diffusion length, characteristic of a bulk spin Hall effect. The distinct thickness dependences therefore indicate that charge-to-spin conversion in the $MoTe_2$/Py bilayers is predominantly interfacial, consistent with a Rashba-Edelstein mechanism, whereas the Pt/Py bilayers are dominated by the conventional bulk spin Hall effect. These results are fully consistent with the SP-FMR measurements and further support the existence of a Rashba interface in the sputtered $MoTe_2$/Py heterostructures.

To quantify the damping-like torque efficiency $\xi_{DL}$, the FM layer thickness was varied while keeping the adjacent layer thickness constant, as described by Eqs. (4.2). The saturation magnetization was assumed to remain approximately constant ($\mu_0 \approx 0.8\ T$) over the investigated thickness range. The values of $1/\xi_{FMR}$ were then plotted as a function of $1/t_{FM}$, and $\xi_{DL}$, was extracted from the corresponding linear fits. Fig. 7(b) shows the linear fit for obtained for the Py(t)/Pt(8) heterostructure, yielding $\xi_{DL} = 0.022$, good agreement with previously reported values for Pt/Py. This value is consistent with the results reported in the literature for Pt/Py[59,61]. The analysis of the $MoTe_2$(3)/Py(t) samples is more complex because the angular asymmetry discussed in Fig. 6 implies that the torque efficiency depends on the in-plane angle. Thus, $\xi_{DL}$ was evaluated at two angles corresponding to the maximum ST-FMR amplitudes, $\phi$ =45º and 225º. Fig. 7(c) shows the linear fit for obtained for the $MoTe_2$(3)/Py(t) heterostructure for $\phi = 45$º, yielding $\xi_{DL} = 0.13$ whereas the analysis at $\phi = 225°$ gives a substantially larger value of $\xi_{DL} = 0.39$ . The strong angular dependence of $\xi_{DL}$ reflects the presence of additional non-conventional damping-like torque components arising from the reduced symmetry of the $MoTe_2$ interface. Similar angular asymmetries have been reported for both single-crystalline and polycrystalline MoTe2-based heterostructures.  and are in agreement with the values found in recent work for single-crystal and polycrystalline $MoTe_2$[31,33,34]. In particular, the value $\xi_{DL} = 0.39$ is remarkably close to the efficiency of $\xi_{DL} = 0.35$ reported from field-free spin-orbit-torque magnetization switching in optimized $MoTe_2$ device[31].

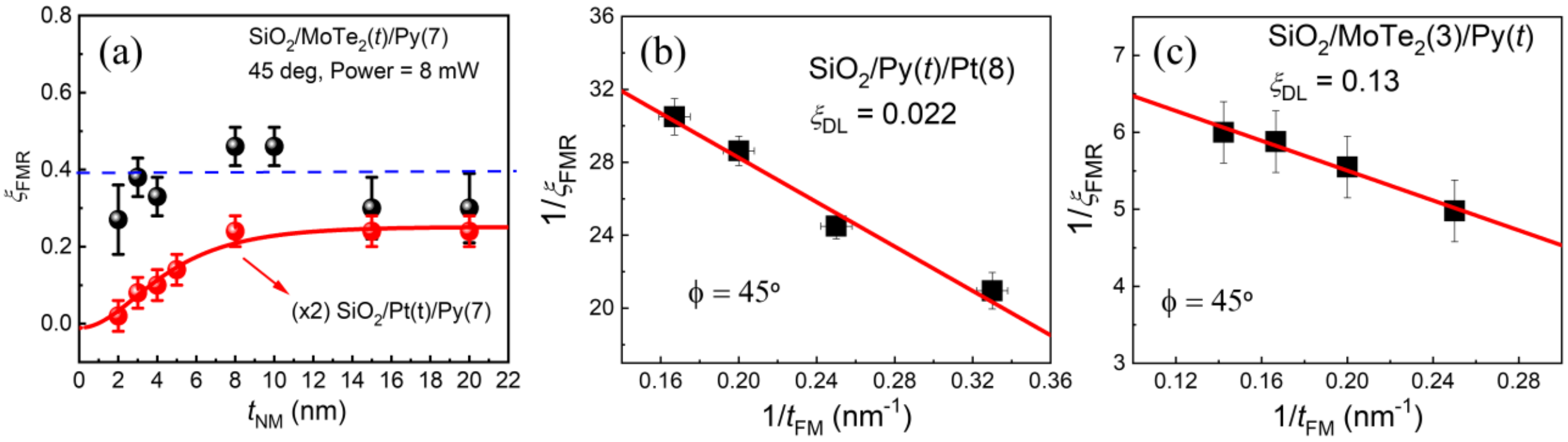

**Figure 7.** (a) Extracted values of $\xi_{FMR}$, as a function of the NM layer thickness, $t_{NM}$, for $SiO_2/MoTe_2(t_{MoTe2})$/Py(7) and $SiO_2/Pt(t_{MoTe2})$/Py(7) heterostructures. The Pt samples (red symbols) exhibit the characteristic asymptotic thickness dependence expected for a bulk SHE, whereas the $MoTe_2$ samples (black symbols) remain nearly independent of thickness, fluctuating around a constant average value, consistent with an interface-dominated charge-to-spin conversion mechanism. The solid red curve is the fit to the bulk SHE model, while the dashed blue line indicates the average value of $\xi_{FMR}$ for the $MoTe_2$ samples. (b) Linear fit of $1/\xi_{FMR}$ versus $1/t_{FM}$ for $SiO_2$/ Py(t)/Pt(8), yielding a damping-like torque efficiency of $\xi_{DL} = 0.022$ calculated for $\phi = 45°$ (c) Linear fit of $1/\xi_{FMR}$ versus $1/t_{FM}$ for $SiO_2/MoTe_2$(3)/Py(t), yielding a significantly larger damping-like torque efficiency of $\xi_{DL} = 0.13$ calculated for $\phi\ = 45°$.

### 2.3 Investigation of ST-FMR and SP-FMR in heterostructures with an Au spacer

To further probe the interfacial origin of the spin-charge conversion, we inserted a 1.5-nm-thick Au spacer layer between the $MoTe_2$ and Py layers and compared the results with those of the corresponding MoTe2/Py bilayer. The Au spacer is expected to suppress the strong Rashba interaction at the MoTe2/Py interface while contributing negligibly to the spin-orbit torques because of its small spin Hall angle. Therefore, if the unconventional out-of-plane torque and the enhanced spin-to-charge conversion originate primarily at the MoTe2/Py interface, both effects should be strongly reduced after insertion of the Au layer.

Figs. 8(a) and 8(b) show the ST-FMR spectra and angular dependence for the $SiO_2/MoTe_2$(3)/Py(7) bilayer, while Figs. 8(c) and 8(d) present the corresponding results for $SiO_2$ /$MoTe_2$(3)/Au(1.5)/Py(7). The angular asymmetry observed in the $MoTe_2$/Py disappears after introducing the Au spacer, indicating that the unconventional out-of-plane torque originates at the $MoTe_2$/Py interface. In addition, a pronounced reduction of the symmetric ST-FMR component is observed. Comparing Figs. 8(b) and 8(d), the symmetric voltage decreases from approximately 0.8 μV to 0.3 μV, corresponding to a reduction of about 63%. This substantial suppression indicates that the interfacial contribution to the spin-charge conversion has been largely eliminated. This interpretation is independently confirmed by the SP-FMR measurements shown in Figs. 8(e) and 8(f). The peak spin-pumping current decreases from approximately 215 nA for the $MoTe_2$/Py bilayer to about 129 nA after insertion of the Au spacer. Moreover, the remaining current is very close to the self-conversion signal measured for the Py(7) reference sample, indicating that the additional contribution associated with the $MoTe_2$/Py interface has been effectively removed. The excellent agreement between the ST-FMR and SP-FMR results provides compelling evidence that both the unconventional torque components and the enhanced spin-to-charge conversion originate predominantly at the $MoTe_2$/Py interface rather than in the bulk of the $MoTe_2$ layer. These findings are fully consistent with the thickness-dependent measurements and the bilateral spin-injection experiments, further establishing the Rashba-Edelstein effect as the dominant conversion mechanism in the sputtered $MoTe_2$/Py heterostructures.

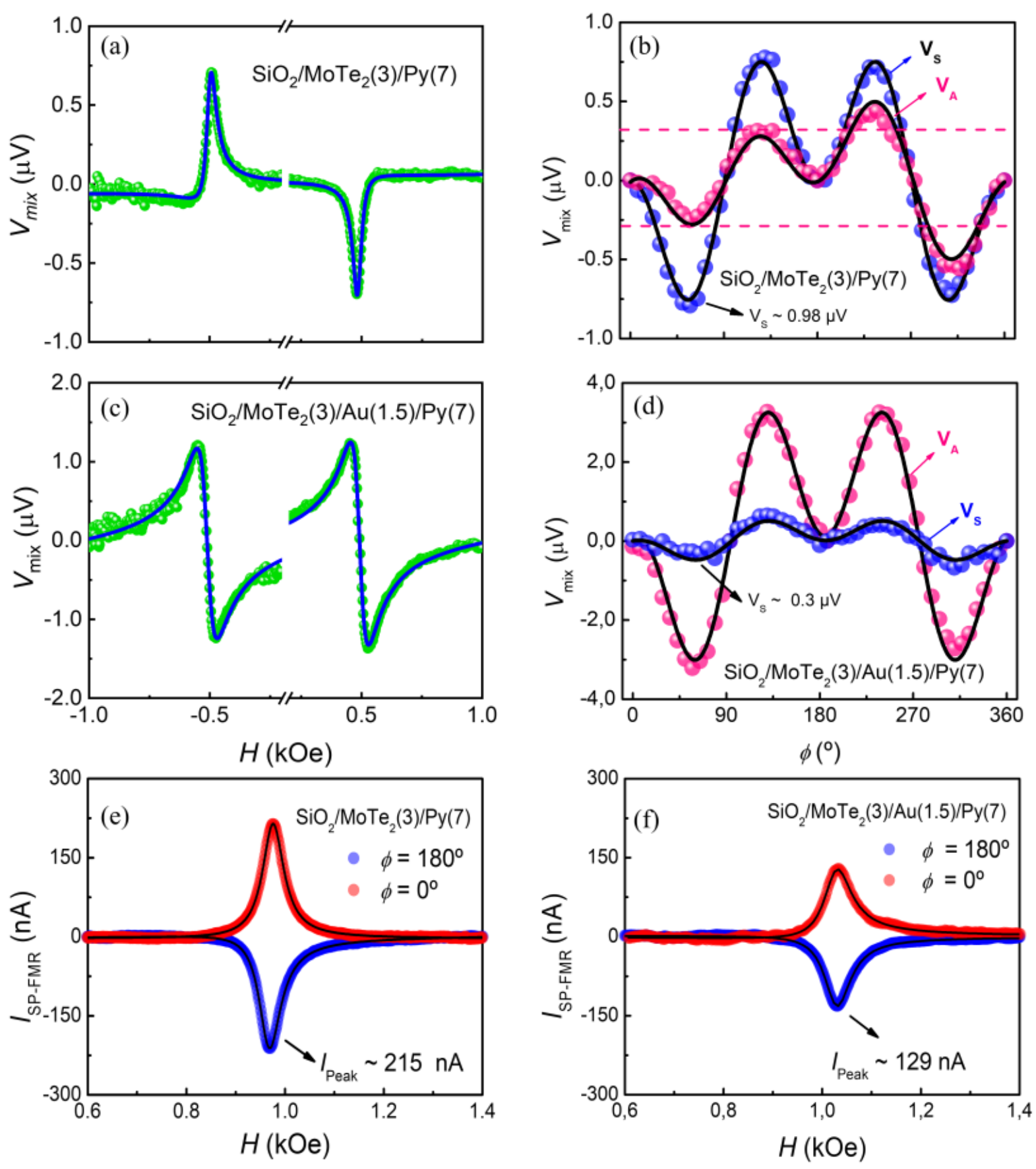

**Figure 8.** Effect of an Au spacer layer on the ST-FMR and SP-FMR response of $MoTe_2$/Py heterostructures. (a) ST-FMR mixing voltage ($V_{\mathrm{mix}}$) and (b) angular dependence of the symmetric $V_S$ and antisymmetric $V_A$ components for the $SiO_2/MoTe_2(3)$/Py(7) heterostructure. (c) ST-FMR mixing voltage ($V_{\mathrm{mix}}$) and (d) angular dependence of the symmetric $V_S$ and antisymmetric $V_A$ components for the $SiO_2/MoTe_2(3)$/Au(1.5)/Py(7) heterostructure. The ST-FMR measurements were performed at a microwave frequency of 6.0 GHz. (e) SP-FMR signals for the SiO2/$MoTe_2(3)$/Py(7) heterostructure, showing a large spin-to-charge conversion generated at the $MoTe_2$/Py interface. (f) SP-FMR signals for the $SiO_2/MoTe_2(3)$/Au(1.5)/Py(7) heterostructure, where the insertion of a 1.5 nm Au spacer strongly suppresses the SP-FMR signal while leaving the ST-FMR response nearly unchanged, demonstrating that the spin-to-charge conversion originates predominantly at the $MoTe_2$/Py interface rather than from the bulk of MoTe2. The rf power was 8 mW for the ST-FMR measurements and 110 mW for the SP-FMR measurements.

## 3. CONCLUSIONS

In this work, we investigated spin-to-charge and charge-to-spin interconversion in sputtered $MoTe_2$/Py heterostructures using ST-FMR, SP-FMR, thickness-dependent measurements, angular analysis, and interface-engineering experiments. Our results consistently demonstrate that the dominant conversion mechanism in these heterostructures originates from interfacial Rashba-Edelstein effects rather than from bulk spin Hall transport. The thickness dependence of the ST-FMR and SP-FMR differs markedly from that observed in Pt reference samples. While Pt exhibits the expected asymptotic behavior associated with bulk spin diffusion and the spin Hall effect, $MoTe_2$ shows nearly thickness-independent conversion efficiencies, supporting an interface-dominated mechanism. Thus, the spin accumulation responsible for the observed torques is primarily generated at the $MoTe_2$/Py interface through Rashba-Edelstein mechanisms. Angular-dependent ST-FMR measurements further revealed unconventional torque symmetries, including out-of-plane torque components absent from the conventional angular dependence of polycrystalline metallic bilayers. These components indicate additional symmetry breaking at the interface, potentially associated with structural inversion asymmetry, interfacial disorder, strain, or Rashba-induced spin textures. The extracted damping-like torque efficiencies reached values as high as $\xi_{DL} \approx 0.39$, demonstrating that sputtered $MoTe_2$ can generate highly efficient spin-orbit torques comparable to or exceeding those reported for conventional heavy metals and high-quality crystalline TMD systems. The interfacial origin of the conversion was further supported by Au spacer-layer experiments. Inserting a thin Au layer between $MoTe_2$ and Py strongly suppressed the angular asymmetry and reduced the enhanced conversion signals, indicating that the unconventional torques and large spin-charge conversion efficiencies are intrinsically linked to the direct $MoTe_2$/Py interface. Moreover, trilayer SP-FMR measurements using opposite spin-current injection directions revealed voltage polarities consistent with the inverse Rashba-Edelstein effect, distinguishing the interfacial conversion mechanism in $MoTe_2$ from the conventional inverse spin Hall effect observed in Pt-based systems. Overall, these results demonstrate that sputtered $MoTe_2$ heterostructures provide an efficient and scalable platform for interface-driven spin-orbit torques. Beyond clarifying the microscopic origin of spin-charge conversion in polycrystalline $MoTe_2$, this work highlights the importance of interface engineering in TMD-based spintronic devices and opens promising perspectives for low-power magnetic memory and logic technologies based on two-dimensional materials.

**Data Availability Statement**

The data that support the findings of this study are available from the corresponding authors upon reasonable request.

**ACKNOWLEDGMENTS**

This research was supported by Conselho Nacional de Desenvolvimento Científico e Tecnológico (CNPq), Coordenação de Aperfeiçoamento de Pessoal de Nível Superior (CAPES) (Grant No. 0041/2022), Financiadora de Estudos e Projetos (FINEP), Fundação de Amparo à Ciência e Tecnologia do Estado de Pernambuco (FACEPE) – (Grant No. BFP-0345 1.05/24), Universidade Federal de Pernambuco, Multiuser Laboratory Facilities of DF-UFPE, Fundação de Amparo à Pesquisa do Estado de Minas Gerais (FAPEMIG) - Rede de Pesquisa em Materiais 2D and Rede de Nanomagnetismo, INCT of Spintronics and Advanced Magnetic Nanostructures (INCT-SpinNanoMag), CNPq 406836/2022-1, and INCT Nanocarbono e Materiais 2D, CNPq 408649/2024-0.

## APPENDIX A - SAMPLE FABRICATION AND CHARACTERIZATION

All thin films were fabricated by DC magnetron sputtering under a base pressure $1.5 \times 10^{-7}$ torr and an argon working pressure $2.7 \times 10^{-3}$ torr. The $MoTe_2$ and Py thin films were grown with applied DC power 10 W and 23 W, respectively. For the heterostructures $SiO_2$/$MoTe_2$/Py, the silicon substrate wafer was the same, but the sample's geometry was relatively different depending on the employed technique. For ST-FMR, the sample's dimensions were 0.150 x 2 mm, while in SP-FMR were 2 x 3 mm. For the electrical voltage detection, conductive silver ink was used to perform the electrical contacts. The Py layers were deposited onto the $MoTe_2$ without breaking the vacuum chamber. The $MoTe_2$ samples were characterized by grazing incidence X-ray diffraction (XRD), Raman spectroscopy, resistivity measurements and atomic force microscopy (AFM) images. The results are present in Fig. A1. The XRD patten is shown in Fig. A1, and it was measured using $K_\alpha$ radiation in a 50 nm thick $MoTe_2$, capped by 1 nm of Au to avoid oxidation. The data reveals the polycrystalline growth of sputtered $MoTe_2$. The Raman spectrum (Fig. A1 (b)) was measured using a 532 nm laser radiation. From the data, we observe that our $MoTe_2$ films are in the 1T' structural phase, characterized by the presence of $A_g$ vibrational modes with narrow peaks, indicating high crystallinity and large grain sizes. The resistivity measurement was acquired using a four-point probe system with Au contacts electrodes, each one 2.5 mm apart. The obtained resistivity ($\rho = 9.69 \cdot 10^{-4} \Omega{\cdot}m$) is compatible with the resistivity of 1T' $MoTe_2$ samples, with minor differences attributed to film thickness, substrate growth and surface roughness. To verify the film quality, we probe the $MoTe_2$ surface using AFM in tapping mode. The surface topography image shows a granular morphology with large grain sizes (Fig. A1(d)). The mean surface roughness is 1.87 nm.

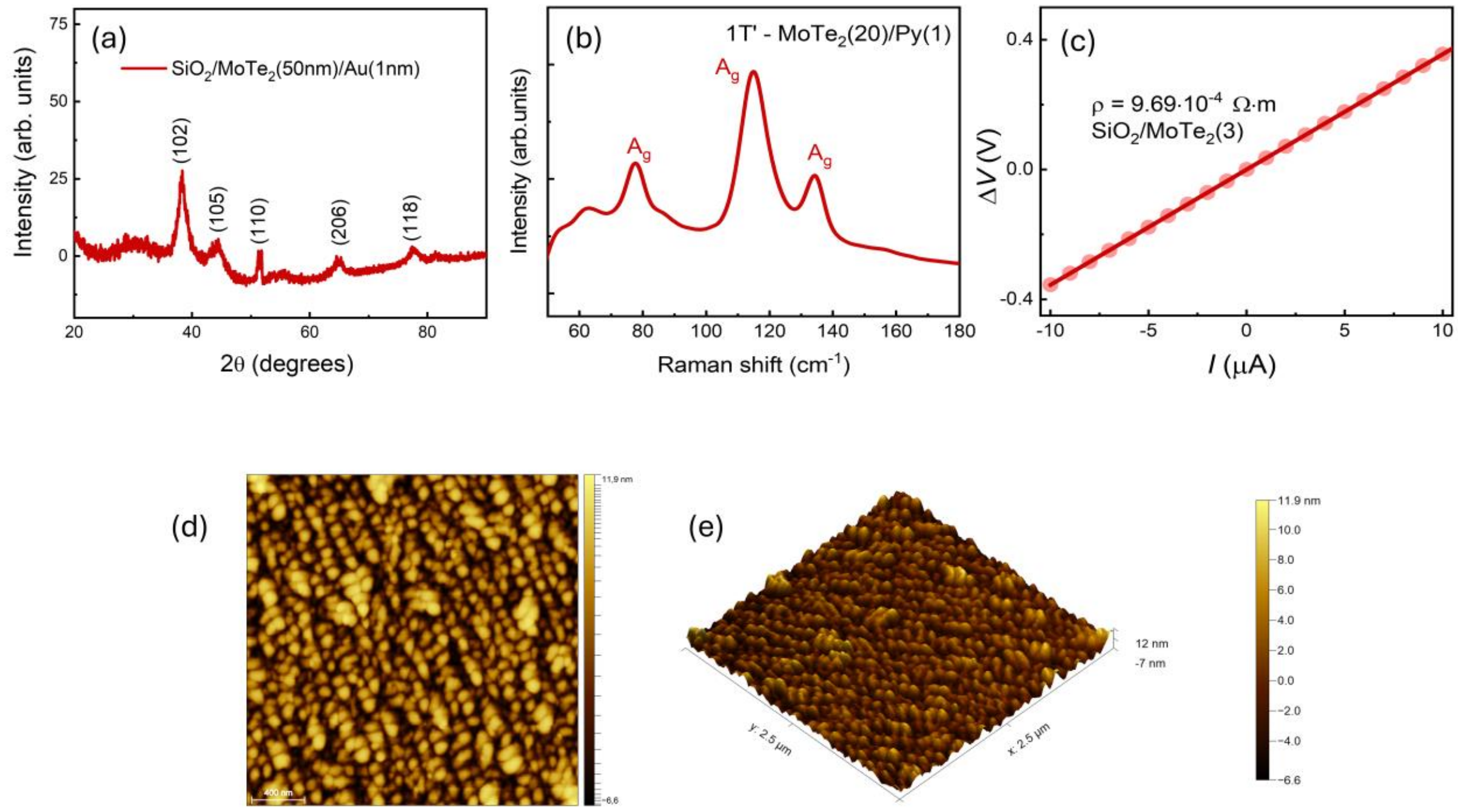

**Figure A1**. (a) X-ray diffraction pattern for $SiO_2/MoTe_2(20)/Au(1)$. (b) Raman spectroscopy data for the sample $SiO_2/MoTe_2(20)/Py(1)$, where the numbers are the layer thickness in nm. The Raman data reveals the presence of 1T' structural phase (c) I x V characteristic curve measured through the four-point probe method. The resistivity from the I x V curve gives compatible resistivity values for the 1T' $MoTe_2$. (d) AFM surface topography image showing a granular morphology with uniform grain distribution. (e) Three-dimensional (3D) AFM surface representation.

## APPENDIX B - EXPERIMENTAL SETUPS AND MEASUREMENTS

The ST-FMR signal is a rectified DC voltage that can be measured directly by a nanovoltmeter or indirectly by lock-in detection. We used lock-in detection because it provides a larger signal-to-noise ratio. The two ends of the NM/FM layer channel are connected to electrodes in a Signal-Ground-Signal coplanar waveguide (CPW), as previously illustrated in Fig. 1(b), having a symmetric configuration and impedance of 50 Ω. The electric connection is made by using conductive silver ink. Before passing through the sample, the rf current is amplitude modulated by a low frequency sinusoidal (7.3 kHz) signal provided by a signal generator. This signal also works as a reference signal for the lock-in. The ST-FMR signal originates from the oscillating AMR caused by the magnetization precession induced by the oscillating torques. Thus, the voltage is rectified because of the combination of oscillating AMR and oscillating rf current. The resulting signal is detected by the lock-in amplifier using the phase locking technique, having symmetric and antisymmetric Lorentzian lineshapes.

**REFERENCES**

(1) K. S. Novoselov, A. K. Geim, S. V. Morozov, D. Jiang, Y. Zhang, S. V. Dubonos, I. V. Grigorieva, and A. A. Firsov. Electric Field Effect in Atomically Thin Carbon Films. *Science*, **306**, Issue 5696 (2004). DOI: 10.1126/science.1102896

(2) S. Z. Butler, S. M. Hollen, L. Cao, et al. Progress, Challenges, and Opportunities in Two-Dimensional Materials Beyond Graphene. *ACS Nano* **7** (4): 2898-2926 (2013). DOI: 10.1021/nn400280c

(3) G. Fiori, F. Bonaccorso, G. Iannaccone, et al. Electronics based on two-dimensional materials. *Nature Nanotech* **9**, 768-779 (2014). DOI: 10.1038/nnano.2014.207

(4) K. S. Kim, J. Kwon, H. Ryu, et al. The future of two-dimensional semiconductors beyond Moore's law. *Nat. Nanotechnol*. **19**, 895-906 (2024). DOI: 10.1038/s41565-024-01695-1

(5) M. C. Lemme, D. Akinwande, C. Huyghebaert, et al. 2D materials for future heterogeneous electronics. *Nat. Commun.* **13**, 1392 (2022). DOI: 10.1038/s41467-022-29001-4

(6) S. Manzeli, D. Ovchinnikov, D. Pasquier, et al. 2D transition metal dichalcogenides. *Nat. Rev. Mater.* **2**, 17033 (2017). DOI: 10.1038/natrevmats.2017.33

(7) Y. Wang, S. Sarkar, H. Yan, and M. Chhowalla. Critical challenges in the development of electronics based on two-dimensional transition metal dichalcogenides. *Nat. Electron.* **7**, 638-645 (2024). DOI: 10.1038/s41928-024-01210-3

(8) D. Voiry, A. Mohite, and M. Chhowalla. Phase engineering of transition metal dichalcogenides. *Chem. Soc. Rev.* **44** (9): 2702-2712 (2015). DOI: 10.1039/c5cs00151j

(9) A. Kuc, T. Heine, and A. Kis. Electronic properties of transition-metal dichalcogenides. *MRS bulletin*, **40**(7), 577-584 (2015). DOI: 10.1557/mrs.2015.143

(10) H. Yang, S. O. Valenzuela, M. Chshiev, et al. Two-dimensional materials prospects for non-volatile spintronic memories. *Nature* **606**, 663-673 (2022). DOI: 10.1038/s41586-022-04768-0

(11) N. Zibouche, A. Kuc, J. Musfeldt, and T. Heine. Transition-metal dichalcogenides for spintronic applications. *Ann. Phys.* **526** 395-401 (2014). DOI: 10.1002/andp.201400137

(12) S. Husain, R. Gupta, A. Kumar, et al. Emergence of spin–orbit torques in 2D transition metal dichalcogenides: A status update. *Appl. Phys. Rev.* **7**, 041312 (2020); DOI: 10.1063/5.0025318

(13) Y. Liu, and Q. Shao. Two-dimensional materials for energy-efficient spin-orbit torque devices. *ACS nano*, **14**(8), 9389-9407 (2020). DOI: 10.1021/acsnano.0c04403

(14) X. Lin, W. Yang, K. L. Wang, and W. Zhao. Two-dimensional spintronics for low-power electronics. *Nat. Electron.* **2**, 274-283 (2019). DOI: 10.1038/s41928-019-0273-7

(15) Y. Deng, X. Zhao, C. Zhu, P. Li, R. Duan, G. Liu, and Z. Liu. $MoTe_2$: semiconductor or semimetal? *ACS nano*, **15**(8), 12465-12474 (2021). DOI: 10.1021/acsnano.1c01816

(16) Z. Wang, D. Gresch, A. A. Soluyanov, et al. $MoTe_2$: A Type-II Weyl Topological Metal. *Phys. Rev. Letts.* **117**, 056805 (2016). DOI: 10.1103/PhysRevLett.117.056805

(17) H.-J. Kim, S.-H. Kang, I. Hamada, and Y.-W. Son. Origins of the structural phase transitions in $MoTe_2$ and $WTe_2$. *Phys. Rev. B* **95**, 180101(R). DOI: 10.1103/PhysRevB.95.180101

(18) M. J. Mleczko, A. C. Yu, C. M. Smyth, et al. Contact engineering high-performance n-type $MoTe_2$ transistors. *Nano letters*, **19**(9), 6352-6362 (2019). DOI: 10.1021/acs.nanolett.9b02497

(19) Q. Shao, G. Yu, Y.-W. Lan, et al. Strong Rashba-Edelstein effect-induced spin-orbit torques in monolayer transition metal dichalcogenide/ferromagnet bilayers. *Nano letters*, **16**(12), 7514-7520 (2016). DOI: 10.1021/acs.nanolett.6b03300

(20) S. Song, D. H. Keum, S. Cho, et al. Room Temperature Semiconductor-Metal Transition of $MoTe_2$ Thin Films Engineered by Strain. *Nano Lett.* **16** (1): 188-193 (2016). DOI: 10.1021/acs.nanolett.5b03481

(21) Y. Wang, J. Xiao, H. Zhu, et al. Structural phase transition in monolayer MoTe2 driven by electrostatic doping. *Nature* **550**, 487-491 (2017). DOI: 10.1038/nature24043

(22) P. Li, W. Wu, Y. Wen, et al. Spin-momentum locking and spin-orbit torques in magnetic nano-heterojunctions composed of Weyl semimetal $WTe_2$. *Nat. Commun.* **9**, 3990 (2018). DOI: 10.1038/s41467-018-06518-1

(23) Q.-F. Yao, J. Cai, W. Tong, et al. Manipulation of the large Rashba spin splitting in polar two-dimensional transition-metal dichalcogenides. *Phys. Rev. B*, **95**(16), 165401 (2017). DOI: 10.1103/PhysRevB.95.165401

(24) A. Manchon, H. C. Koo, J. Nitta, S. M. Frolov, and R. A. Duine. New perspectives for Rashba spin–orbit coupling. *Nature materials*, **14**(9), 871-882 (2015). DOI: 10.1038/nmat4360

(25) M. Yama, M. Matsuo, and T. Kato. Theory of inverse Rashba-Edelstein effect induced by spin pumping into a two-dimensional electron gas. *Phys. Rev. B*, **108**(14), 144430 (2023). DOI: 10.1103/PhysRevB.108.144430

(26) S. Paul, S. Talukdar, R. S. Singh, and S. Saha. Topological phase transition in $MoTe_2$: a review. physica status solidi (RRL)-*Rapid Research Letters*, **17**(6), 2200420 (2023). DOI: 10.1002/pssr.202200420

(27) J. Jiang, Z. K. Liu, Y. Sun, et al. Signature of type-II Weyl semimetal phase in $MoTe_2$. *Nature commun*., **8**(1), 13973 (2017). DOI: 10.1038/ncomms13973

(28) A. Zhang, X. Ma, C. Liu, et al. Topological phase transition between distinct Weyl semimetal states in $MoTe_2$. *Phys. Rev. B,* **100**(20), 201107 (2019). DOI: 10.1103/PhysRevB.100.201107

(29) S. T. Chyczewski, H. Lee, S. Li, M. Eladl, J.-F. Zheng, A. Hoffmann, and W. Zhu. Strong Damping-Like Torques in Wafer-Scale $MoTe_2$ Grown by MOCVD. *ACS Applied Materials & Interfaces*, **17**(14), 21996-22003 (2025). DOI: 10.1021/acsami.4c21247

(30) L. Zhu, D. C. Ralph, and R. A. Buhrman. Spin-orbit torques in heavy-metal-ferromagnet bilayers with varying strengths of interfacial spin-orbit coupling. *Phys. Rev. Lett.*, **122**(7), 077201 (2019). DOI 10.1103/PhysRevLett.122.077201

(31) S. Liang, S. Shi, C.-H. Hsu, et al. Spin-orbit torque magnetization switching in MoTe2/permalloy heterostructures. *Advanced Materials*, **32**(37), 2002799 (2020). DOI: 10.1002/adma.202002799

(32) S. Li, J. Gibbons, S. Chyczewski, et al. Unconventional spin-orbit torques from sputtered MoTe2 films. *Phys. Rev. B* **110**, 024426 2024). DOI: 10.1103/PhysRevB.110.024426

(33) D. MacNeill, G. M. Stiehl, M. H. D. Guimaraes, R. A. Buhrman, J. Park, and D. C. Ralph. Control of spin-orbit torques through crystal symmetry in $WTe_2$/ferromagnet bilayers. *Nature Physics*, **13**(3), 300-305 (2017). DOI: 10.1038/nphys3933

(34) G. M. Stiehl, R. Li, V. Gupta, et al. Layer-dependent spin-orbit torques generated by the centrosymmetric transition metal dichalcogenide $\beta$-$MoTe_2$. *Phys. Rev. B*, **100**(18), 184402 (2019). DOI: 10.1103/PhysRevB.100.184402

(35) W. Choi, N. Choudhary, G. H. Han, J. Park, D. Akinwande, and Y. H. Lee. Recent development of two-dimensional transition metal dichalcogenides and their applications. *Materials Today*, **20**(3), 116-130, (2017). DOI: 10.1016/j.mattod.2016.10.002

(36) S. Xu, J. Zhang, Y. Xia, et al. Single-Site-Directed Unidirectional Epitaxy of Large-Scale 2D Materials. Advanced *Materials* **38**, 13: e20837 (2026). DOI: 10.1002/adma.202520837

(37) C. Bian, Y. Zhao, R. Guzman, et al. Atomically precise synthesis and simultaneous heterostructure integration of 2D transition metal dichalcogenides through nano-confinement. *Nat. Mater.* **25**, 1321-1328 (2026). DOI: 10.1038/s41563-026-02495-9

(38) D. S. Cabeda, A. V. da Silva, and C. Radtke. Influence of the Magnetron Sputtering Power Source Type on the Formation of TMD Thin Films. *Crystal Growth & Design* **26**(4): 1555-1561 (2026). DOI: 10.1021/acs.cgd.5c01221

(39) Y. Tserkovnyak, A. Brataas, and G. E. W. Bauer. Enhanced Gilbert damping in thin ferromagnetic films. *Phys. Rev. Letts.* **88** (11), 117601 (2002). DOI: 10.1103/PhysRevLett.88.117601

(40) Y. Tserkovnyak, A. Brataas, and G. E. W. Bauer. Spin pumping and magnetization dynamics in metallic multilayers. *Phys. Rev. B* **66**, 224403 (2002). DOI: 10.1103/PhysRevB.66.224403

(41) A. Azevedo, L. H. Vilela Leão, R. L. Rodriguez-Suarez, A. B. Oliveira, and S. M. Rezende. dc effect in ferromagnetic resonance: Evidence of the spin-pumping effect? *J. Appl. Phys.* **97**, 10C715 (2005). DOI: 10.1063/1.1855251

(42) E. Saitoh, M. Ueda, H. Miyajima, and G. Tatara. Conversion of spin current into charge current at room temperature: Inverse spin-Hall effect. *Applied physics letters*, **88**(18) (2006). DOI: 10.1063/1.2199473

(43) A. Azevedo, L. H. Vilela-Leão, R. L. Rodríguez-Suárez, A. F. L. Santos, and S. M. Rezende. Spin pumping and anisotropic magnetoresistance voltages in magnetic bilayers: Theory and experiment. *Phys. Rev. B*, **83**, 144402 (2011). DOI:10.1103/PhysRevB.83.144402

(44) J. C. Sánchez, L. Vila, G. Desfonds, et al. Spin-to-charge conversion using Rashba coupling at the interface between non-magnetic materials. *Nat. Commun.* **4**, 2944 (2013). DOI: doi.org/10.1038/ncomms3944

(45) G. Bihlmayer, P. Noël, D. V. Vyalikh, et al. Rashba-like physics in condensed matter. *Nat. Rev. Phys.* **4**, 642-659 (2022). DOI: 10.1038/s42254-022-00490-y

(46) J. L. Costa, E. Santos, G. R. Gallo, G. Rodrigues-Junior, et al. Phase-selective orbital-charge conversion in $MoTe_2$. arXiv:2607.01623 (2026)

(47) Y. S. Gui, L. H. Bai, and C. M. Hu. The physics of spin rectification and its application. *Sci China-Phys Mech Astron*, **56**: 124-141, (2013). DOI: 10.1007/s11433-012-4956-6.

(48) A. Azevedo, R. O. Cunha, F. Estrada, et al. Electrical detection of ferromagnetic resonance in single layers of permalloy: Evidence of magnonic charge pumping. *Phys. Rev. B*, **92**, 024402 (2015). DOI: 10.1103/PhysRevB.92.024402

(49) A. Tsukahara, Y. Ando, Y. Kitamura, et al. Self-induced inverse spin Hall effect in permalloy at room temperature. *Phys. Rev. B*, **89**, 235317 (2014). DOI: 10.1103/PhysRevB.89.235317

(50) L. Bai, P. Hyde, Y. S. Gui, et al. Universal Method for Separating Spin Pumping from Spin Rectification Voltage of Ferromagnetic Resonance. *Phys. Rev. Lett.* **111**, 217602 (2013). DOI: 10.1103/PhysRevLett.111.217602

(51) J. Shen, Z. Feng, P. Xu, D. Hou, Y. Gao, and X. Jin. Spin-to-Charge Conversion in Ag/Bi Bilayer Revisited. *Phys. Rev. Lett.* **126**, 197201 (2021). DOI: 10.1103/PhysRevLett.126.197201.

(52) J. Cheng, B. F. Miao, Z. Liu, et al. Coherent Picture on the Pure Spin Transport between Ag/Bi and Ferromagnets. *Phys. Rev. Lett.* **129**, 097203 (2022). DOI: 10.1103/PhysRevLett.129.097203

(53) J. E. Abrão, E. G. da Silva, G. Rodrigues-Junior, J. B. S. Mendes, and A. Azevedo. Probing the spin-momentum locking on Rashba surfaces via spin current. *ACS Applied Materials & Interfaces*, **17**(9), 13162-13169 (2024). DOI: 10.1021/acsami.4c06090.

(54) E. C. Souza, J. D. M. de Lima, J. L. Costa, G. Rodrigues-Junior, R. O. Cunha, J. B. S. Mendes, and S. M. Rezende. Spin-to-charge conversion mediated by the inverse Rashba–Edelstein effect in polycrystalline thin films of the Weyl semimetal TaP. *Journal of Applied Physics*, **138**(6) (2025). DOI: 10.1063/5.0281343

(55) E. G. da Silva, J. E. Abrão, E. Santos, S. Bedanta, H. F. Ding, J. B. S Mendes, and A. Azevedo. *Appl. Phys. Lett.* **123**, 202402 (2023). DOI: 10.1063/5.0169242.

(56) A. Manchon, J. Železný, I. M. Miron, et al. Current-induced spin-orbit torques in ferromagnetic and antiferromagnetic systems. *Rev. Mod. Phys*. **91**, 035004 (2019). DOI: 10.1103/RevModPhys.91.035004

(57) M. B. Jungfleisch, W. Zhang, J. Sklenar, et al. Interface-driven spin-torque ferromagnetic resonance by Rashba coupling at the interface between nonmagnetic materials. *Phys. Rev. B*, **93**, 224419 (2016). DOI: 10.1103/PhysRevB.93.224419

(58) S Karimeddiny, and D. C. Ralph. Resolving discrepancies in spin-torque ferromagnetic resonance measurements: Lineshape versus linewidth analyses. *Phys. Rev. Applied*, **15**, 064017 (2021). DOI: 10.1103/PhysRevApplied.15.064017.

(59) L. Liu, T. Moriyama, D. C. Ralph, and R. A. Buhrman. Spin-Torque Ferromagnetic Resonance Induced by the Spin Hall Effect. *Phys. Rev. Lett.*, **106**, 036601 (2011). DOI: 10.1103/PhysRevLett.106.036601

(60) M.-H. Nguyen, and C.-F. Pai. Spin-orbit torque characterization in a nutshell. *APL Mater.*, **9**, 030902 (2021). DOI: 10.1063/5.0041123.

(61) J. L. Costa, E. Santos, A. Y. M. Tani, et al. Investigating spin and orbital effects via spin-torque ferromagnetic resonance. *J. Appl. Phys.*, **139**, 243904 (2026). DOI: 10.1063/5.0320961.

(62) N. Soya, S. Yoshikawa, T. Katase, and K. Ando. Spin-torque ferromagnetic resonance in $SrTiO_3$-based systems: impact of out-of-phase Oersted field torque. *npj Spintronics* **3**, 38 (2025). DOI: 10.1038/s44306-025-00102-2.